# Compact Analytical Model of Dual Material Gate Tunneling Field Effect Transistor using Interband Tunneling and Channel Transport

Rajat Vishnoi and M. Jagadesh Kumar, *Senior Member, IEEE*

*Abstract*— In this paper we have developed a two dimensional (2D) analytical model for surface potential and drain current for a long channel Dual Material Gate (DMG) Silicon-on-Insulator (SOI) Tunneling Field Effect Transistor (TFET). This model includes the effect of drain voltage, gate metal work function, oxide thickness and silicon film thickness, without assuming a fully depleted channel. The proposed model also includes the effect of charge accumulation at the interface of the two gates and the variation in the tunneling volume with the applied gate voltage. The accuracy of the model is tested using two-dimensional numerical simulations. In comparison to the conventional TFET, the proposed model predicts that a DMGTFET provides a higher ON-state current ($I_{ON}$), a better ON-state to OFF-state current ($I_{ON}/I_{OFF}$) ratio and a better sub-threshold slope (SS).

## I. INTRODUCTION

With the dimensions of CMOS transistors reaching sub-100 nm ranges, traditional MOSFETs pose several problems like high leakage currents in OFF-state, high subthreshold slope (SS), drain induced barrier lowering (DIBL) and numerous other short channel effects (SCE). These problems lead to a higher power consumption posing difficulties in scaling down the supply voltage. Hence, as an alternative to MOSFETs, TFETs have been widely studied [1-6]. TFETs exhibit SS below 60 mV/decade, low off state leakage currents and diminished short channel effects due to the built-in tunneling barrier. However, TFETs have limitations of their own. Their ON-state current ($I_{ON}$) is lower than the ITRS requirement [3, 7]. The SS and the leakage currents of TFETs are lower than that of a MOSFET, however they need further improvements. TFETs also suffer from delayed saturation which can be detrimental in analog applications [8]. Sometimes there can also be strong DIBL effects in TFETs [9]. In an attempt to solve these problems, a TFET with DMG (Dual Material Gate) was proposed [10] (Fig. 1) in which the tunneling gate has a work function lower than that of the auxiliary gate for an n-channel TFET and vice versa for a p-channel TFET. DMGTFETs have been demonstrated to have a higher $I_{ON}$ due to the increased tunneling by a metal of lower work function [3, 10]. The OFF-state current ($I_{OFF}$) is reduced because of the presence of a minimum in the surface potential and a negative electric field in the channel, as a result of which we get a better $I_{ON}/I_{OFF}$ ratio and a better SS [3, 10, 11]. Also the drain saturation voltage is reduced [10, 11]. The DMGTFET can be fabricated using a self-aligned symmetric spacer process [12] and has been explored extensively in literature [13-15].

Although, different aspects of DMGTFETs have been studied using TCAD numerical simulations, a compact analytical model will be useful for circuit design and will provide a better insight into the functioning of the device. A number of analytical models have been reported for the SMGTFET [16-22]. However, accurate analytical models for the DMGTFET still need to be developed. The objective of this work is, therefore, to develop a compact analytical model for the DMGTFET using a pseudo-2D approach [23] for solving the Poisson equation together with a combination of interband tunneling and channel transport [22].

In this paper, we first model the surface potential along the length of the p-channel DMGTFET highlighting the step introduced in the potential curve at the interface of the two gates and then using this potential distribution we derive the drain current using Kane's model [24] for tunneling. The potential modeling is done by using a pseudo-2D model [23] for solving the Poisson equation in the silicon channel. To model the drain current, we find the shortest tunneling path using the potential distribution at the source end and then use the average electric field along the shortest tunneling path in the Kane's model as done in [22]. Later we demonstrate, using the surface potential and drain current models, the advantages provided by the DMGTFET over a conventional TFET with SMG (Single Material Gate). These models can serve as important tools for designing DMGTFETs and understanding their behavior. The effect of varying device parameters can be easily studied using these models.

To validate our model we first reproduce the experimental results given in Fig 6 (a) of [22] and set the tunneling parameters for the TCAD simulations [25]. We then compare the potential profile as well as the current characteristics with numerical simulations.

The authors are with the Department of Electrical Engineering, Indian Institute of Technology, Delhi, 110 016 India
(e-mail: vishnoir@gmail.com; mamidala@ee.iitd.ac.in).

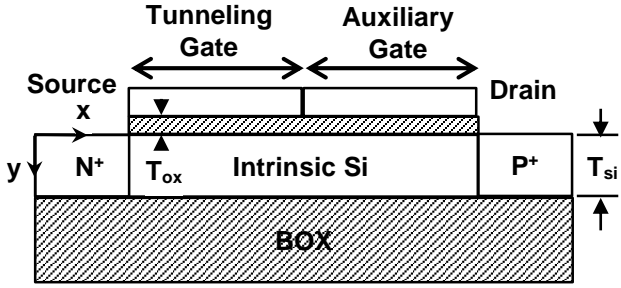

Fig1. Schematic view of the P-channel DMG TFET used in our study.

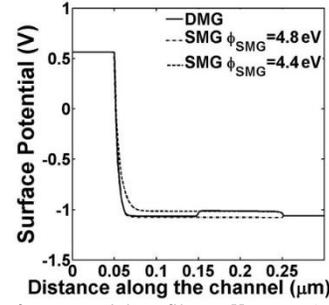

Fig 3. Simulated surface potential profiles at $V_{GS} = -1.5$ V and $V_{DS} = -0.5$ V of the DMGTFET compared with that of the two SMGTFETs having gate work functions $\Phi_{SMG} = 4.4$ eV and 4.8 eV.

## II. MODEL DERIVATION

Fig. 1 shows the schematic view of our device structure with the following parameters [22]: channel length = 200 nm, length of source and drain regions = 50 nm each, silicon film thickness ($T_{si}$) = 10 nm, gate oxide thickness ($T_{ox}$) = 2 nm, source/drain doping concentration = $10^{21}$ cm$^{-3}$ and the body doping $N_S = 10^{15}$ cm$^{-3}$. The two gates are tunneling gate at the source side and auxiliary gate at the drain side with work functions $\Phi_{tunnel} = 4.8$ eV and $\Phi_{aux} = 4.4$ eV, respectively, as suggested in [10]. As a general case, we first assume $L_t$ and $L_a$ (length of tunneling gate and auxiliary gate, respectively) to be 100 nm each and later extend our model for $L_t$ = 20 nm and $L_a$ = 180 nm [10].

Fig. 2 shows the band diagram and surface potential of the DMGTFET along the channel length. One key observation here is that as the current in a TFET is low, the potential drop along the channel is negligible in the regions shown by the solid arrows in the figure and hence it can be assumed to be constant [22]. We can, therefore, infer that the channel is not depleted in these regions. In Fig. 3, we have compared the surface potential of DMGTFET with that of SMGTFET with gate work functions $\Phi_{SMG} = 4.4$ eV and 4.8 eV. Here, we observe that in the DMGTFET, the two non-depleted regions in the channel, one under each gate, has a potential value equal to that in a SMGTFET of the corresponding gate work function. We define the values of these constant potential regions under the tunneling and the auxiliary gates as $\psi_{Ct}$ and $\psi_{Ca}$, respectively.

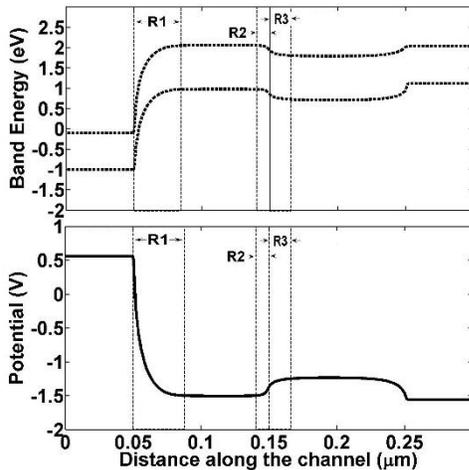

Fig 2. Simulated band diagram (upper curve) and surface potential (lower curve) of the DMGTFET at $V_{GS} = -1.5$ V and $V_{DS} = -1.0$ V. The depletion regions are marked by regions R1, R2 and R3 and the non-depleted regions are shown by solid arrows.

Apart from the tunneling region at the source end, TFETs behave like regular MOSFETs particularly in the channel region since the mechanism of channel formation and charge transport in the channel is the same. Hence, the value of $\psi_{Ct}$ is given by

$$\psi_{Ct} = V_{DS} + \psi_{Bt} \quad \text{if } |V_{DS}| \leq |V_{GS} - V_{tha}| \quad (1)$$
$$\psi_{Ct} = V_{GS} - V_{tha} + \psi_{Bt} \quad \text{if } |V_{DS}| \geq |V_{GS} - V_{tha}| \quad (2)$$

where $\psi_{Bt}$ is the channel's built-in potential under the tunneling gate which is the sum of the amount of band bending caused by the applied $V_{GS}$ and the drop across the buried oxide, and $V_{tha}$ is the threshold voltage for a MOSFET with a gate work function of $\Phi = \Phi_{aux}$. As a general case if $\Phi_{aux} > \Phi_{tunnel}$ then in the expression for $\psi_{Ct}$, $V_{tha}$ would be replaced by $V_{tht}$ which is the threshold voltage for a MOSFET with gate of $\Phi = \Phi_{tunnel}$. This happens because for a p-channel TFET if $\Phi_{aux} < \Phi_{tunnel}$ then $|V_{tha}| > |V_{tht}|$. Therefore, with increasing $V_{DS}$ saturation happens in the auxiliary channel first and the entire channel potential gets saturated. However, if $\Phi_{aux} > \Phi_{tunnel}$ then $|V_{tha}| < |V_{tht}|$ and the channel under the tunneling gate gets saturated first and the auxiliary channel potential is still under the influence of the drain voltage. Hence, it can be demonstrated that by using DMG with $\Phi_{aux} < \Phi_{tunnel}$, we can achieve a lower $V_{Dsat}$ than in an SMGTFET with $\Phi_{SMG} = \Phi_{tunnel}$ which solves the problem of delayed saturation in a TFET. The value of $\psi_{Ca}$ is given by

$$\psi_{Ca} = V_{DS} + \psi_{Ba} \quad \text{if } |V_{DS}| \leq |V_{GS} - V_{tha}| \quad (3)$$
$$\psi_{Ca} = V_{GS} - V_{tha} + \psi_{Ba} \quad \text{if } |V_{DS}| \geq |V_{GS} - V_{tha}| \quad (4)$$

where $\psi_{Ba}$ is the channel built-in potential under the auxiliary gate which is the sum of the amount of band bending caused by the applied $V_{GS}$ and the drop across the buried oxide. Also,

$$\psi_{Source} = V_S + V_{bi} \quad (5)$$
$$\psi_{Drain} = V_{DS} + V_S \quad (6)$$

where $V_{bi}$ is the built-in potential at the source-body junction and $V_S$ is at ground potential.

There are three depletion regions in the channel as shown in Fig. 2 (R1, R2 and R3) and we need to model the surface potential in these regions. Using the pseudo-2D approach [23] and $\psi_{Source}$ and $\psi_{Ct}$ as boundary values, the potential at the source end i.e. region R1 is modeled in [22] and is given by

$$\psi_{s1}(x) = (\psi_{Ct} - \psi_{Gt}) \times cosh\left(\frac{(x-L_1)}{L_d}\right) + \psi_{Gt} \quad (7)$$

where $L_1$ is the length of region R1 and can be evaluated by setting $\psi_{s1}(0) = \psi_{Source}$, $\psi_{Gt}$ is the electrostatic potential of the tunneling gate and is equal to $V_{GS} - V_{FBt}$ where $V_{FBt}$ is the flat band voltage for the tunneling gate.

For regions R2 and R3, the potential profile can be modeled by solving the 2D Poisson equation (8) in these regions separately using the pseudo 2D approach [23]:

$$\frac{\partial^2 \psi(x,y)}{\partial^2 x} + \frac{\partial^2 \psi(x,y)}{\partial^2 y} = \frac{qN_T}{\varepsilon_{Si}} \quad (8)$$

where $N_T$ is the net negative charge concentration in regions R2 and R3. One important point here is that the value of $N_T$ will be different for regions R2 and R3. The channel at the boundary between R2 and R3 behaves like an $n^+ - n$ junction. Hence, there will be complete depletion in R2 and the mobile charges of region R2 will move into region R3. We assume this charge in region R3 to be $n$ (/cm³). Hence $N_T$ in region R2 will be equal to the background doping $N_S$ of the silicon film and in region R3, it will be equal to $N_S + n$. Following the approach in [23], with one change in the vertical boundary conditions (equation (3) of [23]), that is, the electric field at the silicon film and buried oxide interface should be taken to be zero, for each region $R_j$ we get the following general solution for the surface potential of a single gate DMGTFET assuming full depletion (in vertical direction) in the silicon film:

$$\psi_{sj}(x) = C_j exp\left(\frac{x-L_j}{L_d}\right) + D_j exp\left(\frac{-(x-L_j)}{L_d}\right) + \psi_{Gj} - \frac{qN_{Tj}T_{Si}}{C_{ox}} \quad (9)$$

$$L_d = \sqrt{\frac{\varepsilon_{si}}{\varepsilon_{ox}} T_{si} T_{ox}} \quad (10)$$

where $C_j$ and $D_j$ are unknown coefficients, $L_j$ is the length of the $j^{th}$ region and $L_d$ is the characteristic length. We will have two equations of the form of Eq. (9), one each for regions R2 and R3. Hence, we will have six unknown parameters ($C_2$, $D_2$, $L_2$, $C_3$, $D_3$, $L_3$). These unknowns can be determined by the following six boundary conditions defining $x = 0$ at the junction of the two gates.

The value of $\psi_S$ at $x = -L_2$ and $x = L_3$, respectively, are given by

$$\psi_{S2}(-L_2) = \psi_{Ct} \quad (11)$$
$$\psi_{S3}(L_3) = \psi_{Ca} \quad (12)$$

The electric field at $x = -L_2$ and at $x = L_3$ is zero.

$$\frac{\partial \psi_{s2}(-L_2)}{\partial x} = 0 \quad (13)$$
$$\frac{\partial \psi_{s3}(L_3)}{\partial x} = 0 \quad (14)$$

The surface potential is continuous at $x = 0$.

$$\psi_{S2}(0) = \psi_{S3}(0) \quad (15)$$

The electric field is continuous at $x = 0$.

$$\frac{\partial \psi_{s2}(0)}{\partial x} = \frac{\partial \psi_{s3}(0)}{\partial x} \quad (16)$$

From (11) and (13), we get

$$C_2 = D_2 = \frac{\psi_{Ct} - \psi_{Gt} + \frac{qN_S T_{si}}{C_{ox}}}{2} \quad (17)$$

From (12) and (14), we get

$$C_3 = D_3 = \frac{\psi_{Ca} - \psi_{Ga} + \frac{q(N_S+n)T_{si}}{C_{ox}}}{2} \quad (18)$$

From (15) we get:

$$\frac{\psi_{Ct} - \psi_{Gt} + \frac{qN_S T_{si}}{C_{ox}}}{2} \times \left(e^{\frac{L_2}{L_d}} + e^{\frac{-L_2}{L_d}}\right) + \psi_{Gt} - \frac{qN_S T_{si}}{C_{ox}} =$$
$$\frac{\psi_{Ca} - \psi_{Ga} + \frac{q(N_S+n)T_{si}}{C_{ox}}}{2} \times \left(e^{\frac{L_3}{L_d}} + e^{\frac{-L_3}{L_d}}\right) + \psi_{Ga} - \frac{q(N_S+n)T_{si}}{C_{ox}} \quad (19)$$

From (16) we get:

$$\frac{\psi_{Ct} - \psi_{Gt} + \frac{qN_S T_{si}}{C_{ox}}}{2} \times \left(e^{\frac{L_2}{L_d}} - e^{\frac{-L_2}{L_d}}\right)$$
$$= \frac{\psi_{Ca} - \psi_{Ga} + \frac{q(N_S+n)T_{si}}{C_{ox}}}{2} \times \left(e^{\frac{-L_3}{L_d}} - e^{\frac{L_3}{L_d}}\right) \quad (20)$$

The value of $n$ in region R3 can be calculated from the following condition:

$$nL_3 = n_{ch2}L_2 \quad (21)$$

where $n_{ch2}$ is the inversion charge concentration under the tunneling gate at the given gate voltage. There are non-linear terms in (19) and (20) and therefore, they have to be solved numerically. Simultaneously solving (19), (20) and (21) gives us the values of $L_2$ and $L_3$. Since $L_2$ and $L_3$ are of the same order and the inversion charge concentration is typically $10^{19}$ cm⁻³, $n$ can be taken to be equal to $10^{19}$ cm⁻³. With this assumption, $L_2$ and $L_3$ can be found by solving only Eqs. (19) and (20).

As shown in Fig. 2, in region R1, the slope of the surface potential decreases along the channel length and hence the shortest tunneling length $L_T$ must lie between the source and the point where the potential falls by $E_G/q$ and can be written as [22]:

$$L_T = x(\psi_{source}) - x(\psi_{source} - E_G/q) \quad (22)$$

$$x(\psi_s) = L_d \cosh^{-1} \frac{(\psi_s - \psi_{Gt})}{(\psi_{Ct} - \psi_{Gt})} \quad (23)$$

The average electric field $E_{avg}$ along the shortest tunneling path is given by

$$E_{avg} = \frac{E_G}{qL_T} \quad (24)$$

Substituting $E_{avg}$ into the Kane's model equation for band to band tunneling current [24], the drain current can be calculated as

$$I_D = A_K \times L_T \times E_{avg}^2 \times exp\left(\frac{-B_K}{E_{avg}}\right) \quad (25)$$

where $A_K$ and $B_K$ are tunneling parameters which depend on $E_G$ and the effective mass of the carriers. The term $A_K \times L_T$

incorporates the tunneling volume. Here, we have included the change in tunneling volume with $V_{GS}$ by assuming the tunneling volume to be proportional to the shortest tunneling length $L_T$. The other dimensions of the tunneling volume will be constant for a given silicon film thickness and are part of $A_K$.

Most of the studies on DMGTFETs till date [10, 11] suggest that tunneling gate length $L_t$ should be much smaller than the auxiliary gate length $L_a$. As stated earlier, [10] suggests that for a device of length 50 nm the optimal $L_t$ = 20 nm. In such a structure, it is possible that regions R1 and R2 may merge into each other at low values of $V_{GS}$ giving a full depletion region under the tunneling gate. Hence, in this case we will have to modify our approach for surface potential modeling. We now have only two depletion regions: region R1 which is the entire region under the tunneling gate and region R3 which is the same as earlier. Therefore, we will have two equations similar to (9) once again (one each for R1 and R3, respectively) and hence six unknowns ($C_1$, $D_1$, $L_1$, $C_3$, $D_3$, $L_3$). We will again use six boundary conditions as earlier but with one change (defining $x = 0$ as the junction of the two gates).

The tunneling gate length $L_t$ is the length of region R1 now and not $L_1$. Since regions R1 and R2 have merged, we will get a point of minimum in the surface potential at $x = -L_1$. As a result, the condition given by equation (11) will be different now. The value of $\psi_{s1}$ at $x = -L_t$ is given by

$$\psi_{s1}(-L_t) = V_{bi} \quad (26)$$

All the other boundary conditions given by equations (12)-(16) will remain the same but the variables and constants of region R2 will be replaced by those of region R1. (i.e. $\psi_{s2}$ will become $\psi_{s1}$ and so on). Finally, solving in a similar manner as earlier gives us the following equations.

$$C_3 = D_3 = \frac{\psi_{Ca} - \psi_{Ga} + \frac{q(N_S+n)T_{si}}{C_{ox}}}{2} \quad (27)$$

$$C_1 = D_1 = \frac{V_{bi} - \psi_{Gt} + \frac{qN_S T_{si}}{C_{ox}}}{2\cosh\left(\frac{-L_t+L_1}{L_d}\right)} \quad (28)$$

$$\frac{V_{bi} - \psi_{Gt} + \frac{qN_S T_{si}}{C_{ox}}}{2\cosh\left(\frac{-L_t+L_1}{L_d}\right)} \times \left(e^{\frac{L_1}{L_d}} + e^{\frac{-L_1}{L_d}}\right) + \psi_{Gt} - \frac{qN_S T_{si}}{C_{ox}} =$$
$$\frac{\psi_{Ca} - \psi_{Ga} + \frac{q(N_S+n)T_{si}}{C_{ox}}}{2} \times \left(e^{\frac{L_3}{L_d}} + e^{\frac{-L_3}{L_d}}\right) + \psi_{Ga} - \frac{q(N_S+n)T_{si}}{C_{ox}} \quad (29)$$

$$\frac{V_{bi} - \psi_{Gt} + \frac{qN_S T_{si}}{C_{ox}}}{2\cosh\left(\frac{-L_t+L_1}{L_d}\right)} \times \left(e^{\frac{L_1}{L_d}} - e^{\frac{-L_1}{L_d}}\right) = \frac{\psi_{Ca} - \psi_{Ga} + \frac{q(N_S+n)T_{si}}{C_{ox}}}{2} \times \left(e^{\frac{-L_3}{L_d}} - e^{\frac{L_3}{L_d}}\right)$$
$$(30)$$

Solving equations (28)-(30) simultaneously gives the values of $C_1$, $L_1$ and $L_3$.

Therefore, for structures with small tunneling gate length $L_t$, we will first solve for $L_1$, $L_2$ and $L_3$ using equations (17)-(20) and check whether $L_1+L_2$ is greater than $L_t$ (which will happen at low gate voltages) in which case regions R1 and R2 will merge into each other and we use equations (26)-(30) for modeling the surface potential, otherwise we do it with equations (17)-(20). The value of $I_D$ can then be calculated using equations (22)-(25).

### III. MODEL VALIDATION

The models for surface potential and drain current proposed in the previous section are verified using two-dimensional simulation. In our simulations [25], we have used the models for concentration dependent mobility, electric field dependent mobility, SRH recombination, auger recombination, band gap narrowing and Kane's band to band tunneling. The device structure of Fig. 1 is simulated with the tunneling parameters $A_{Kane}$ and $B_{Kane}$ [25]. The values of $A_{Kane}(= 4.0 \times 10^{19} \text{ eV}^{1/2}/\text{cm-s-V}^2)$ and $B_{Kane}(= 41 \text{ MV/cm-}eV^{3/2})$ are extracted by accurately reproducing the experimental results given in Fig. 6(a) of [22] as shown in Fig 4. The surface potential is plotted in Fig. 5 for different values of $V_{GS}$ and $V_{DS}$ by taking a horizontal cut line under the gate and is compared with the model given by equations (17)-(20). In Fig. 6 and Fig. 7, we compare the simulated $\log(I_D) - V_{GS}$ and $\log(I_D) - V_{DS}$ curves, respectively, with those predicted by the model given by Eqs. (22)-(25). Here $V_{tha}$ is taken to be $-0.9$ V and for simplicity $\psi_{Bt}$ is taken to be constant [22] i.e. $-0.7$ V. In Fig. 8, the surface potential curves are plotted for a DMGTFET with $L_t$ = 20 nm and $L_a$ = 180 nm at low gate voltages as given by equations (26)-(30) and compared with simulation results. The potential model plotted in Fig. 5 is in good agreement with the simulation results. So is the case with current model plotted in Fig. 6 except for small values of $V_{GS}$. In Fig. 7, the model predicts the drain current accurately for $V_{DS}$ above -0.5 V. For $V_{DS}$ below -0.5 V, our model loses its accuracy because at low $V_{DS}$ our model underestimates the average electric field in the tunneling region. It can be observed from Fig. 8 that the model given by equation (26)-(30) is tracking the potential with good accuracy. It predicts the occurrence of a minimum in the surface potential under the tunneling gate. This minimum has a value greater than $\psi_{Ct}$ and thus causes an increase in $L_t$ at low gate voltages leading to a reduction in OFF-state current giving us a better SS and $I_{ON}/I_{OFF}$ ratio.

### IV. PERFORMANCE PREDICTION AND DISCUSSIONS

In this section we will demonstrate how our model predicts the behavior of DMGTFETs and the benefits offered by them over the conventional SMGTFETs.

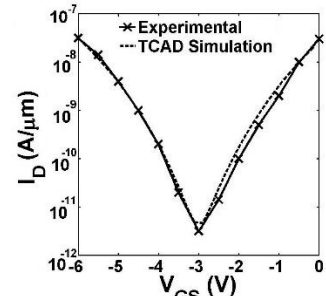

Fig 4. Reproduction of experimental results in Fig. 6(a) of [22] using TCAD simulations for extracting tunneling parameters $A_{Kane}$ and $B_{Kane}$.

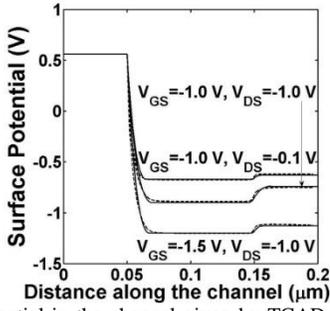

Fig 5. Surface potential in the channel given by TCAD simulations (dashed lines) and our model (solid lines) for three biasing cases.

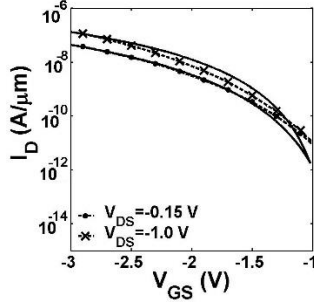

Fig 6. $\log(I_D) - V_{GS}$ curves for the DMGTFET obtained by TCAD simulations (dashed lines) and our model (solid lines) for two values of $V_{DS}$.

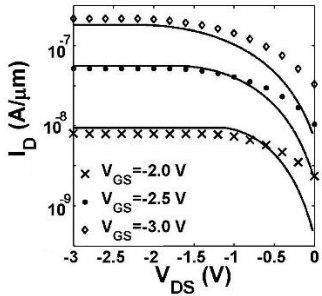

Fig 7. $\log(I_D) - V_{DS}$ curves for the DMGTFET obtained by TCAD simulations (dashed lines) and our model (solid lines) for three values of $V_{GS}$.

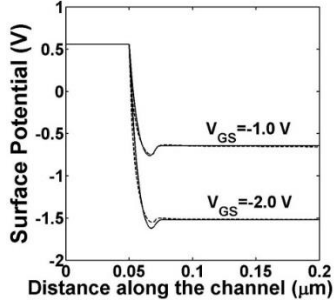

Fig 8. Surface potential profile in the channel obtained from simulations (dashed lines) with $L_t = 20$ nm and $L_a = 180$ nm and model (solid lines) for $V_{DS} = -1.0$ V and two low values of $V_{GS}$.

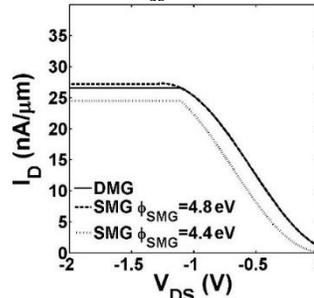

Fig 9. Model predicted drain current versus drain voltage for the DMGTFET and SMGTFETs with $\Phi_{SMG} = 4.4$ eV and 4.8 eV, channel length 200 nm, $T_{si} = 10$ nm, $T_{ox} = 2$ nm, for $V_{GS} = -2.0$ V.

Fig. 9 shows the comparison between the $I_D - V_{DS}$ curves of our DMGTFET structure as predicted by our model and of a SMGTFETs with $\Phi_{SMG} = 4.4$ eV and 4.8 eV as given by the model in [22]. The figure clearly indicates that due to the use of an auxiliary gate of smaller work function, the DMGTFET saturates at a drain voltage which is 0.2 V smaller in magnitude than a SMGTFET with $\Phi_{SMG} = 4.8$ eV. Also, it can be seen that the DMGTFET has a higher $I_{ON}$ as compared to a SMGTFET with $\Phi_{SMG} = 4.4$ eV.

In Fig. 10, we have shown the variation in the shortest tunneling length $L_T$ versus $V_{GS}$ for a DMGTFET with $L_t = 20$ nm and $L_a = 180$ nm as predicted by our models and that for two SMGTFETs of work functions $\Phi_{SMG} = 4.8$ eV and 4.4 eV given by the models in [22]. It can be clearly observed from the figure that $L_T$ for a DMGTFET is larger at OFF-state voltages as compared to a SMGTFET with $\Phi_{SMG} = 4.8$ eV and similar at ON-state voltages. Thus, it can be predicted that the DMGTFET will have a better $I_{ON}/I_{OFF}$ ratio and a better SS as compared to a SMGTFET with $\Phi_{SMG} = 4.8$ eV. It will also have a larger $I_{ON}$ as compared to a SMGTFET with $\Phi_{SMG} = 4.4$ eV which otherwise has a better $I_{ON}/I_{OFF}$ ratio and SS. Therefore, using a DMGTFET we get benefits of both the SMGTFETs (with work functions $\Phi_{SMG} = 4.8$ eV and 4.4 eV) simultaneously. Also Fig.11, where we have plotted the electric field along the channel of the DMGTFET as given by our model, shows that there will be a negative electric field in the middle of the channel which will decelerate the carriers and reduce the drain current. This effect will be more profound during the OFF-state when the amount of carriers will be less as compared to the ON-state and hence it will further improve the $I_{ON}/I_{OFF}$ ratio.

Fig.12 shows the model results for $\log(I_D) - V_{GS}$ curves of a DMGTFET with $L_t = 20$ nm and $L_a = 180$ nm as given by our models and that for a SMGTFET with $\Phi_{SMG} = 4.8$ eV. Since the drain current of DMGTFET has a sharper rise than that of SMGTFET, it is evident that the DMGTFET has a better average subthreshold swing ($SS_{AVG}$) than a SMGTFET, where $SS_{AVG}$ is given by [10]

$$SS_{AVG} = \frac{V_T - V_{OFF}}{\log[I_{DS}(V_T)] - \log[I_{DS}(V_{OFF})]} \quad (31)$$

where $V_T$ is the value of $V_{GS}$ at which $I_D$ is equal to $10^{-8}$ A/μm and $V_{OFF}$ is the gate voltage at which $I_D$ starts to take off.

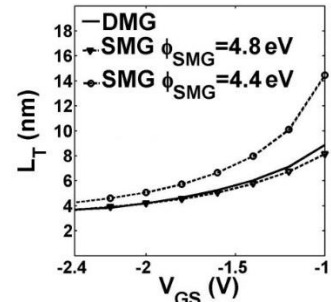

Fig 10. Shortest tunneling length versus gate voltage for the DMGTFET (solid line) with $L_t = 20$ nm and $L_a = 180$ nm and SMGTFETs (dashed line) with work functions $\Phi_{SMG} = 4.8$ eV and 4.4 eV, channel length 200 nm, $T_{si} = 10$ nm, $T_{ox} = 2$ nm, obtained by our model at $V_{DS} = -1.0$ V.

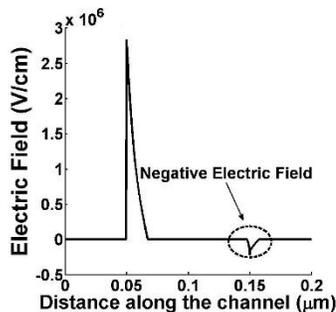

Fig 11. Electric field along the surface of the DMGTFET for $V_{GS} = -1.5$ V, $V_{DS} = -1.0$ V obtained by differentiating the surface potential shown in Fig.5.

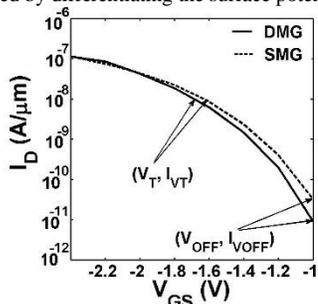

Fig 12. $\log(I_D) - V_{GS}$ for the DMGTFET (solid line) with $L_t = 20$ nm and $L_a = 180$ nm and the SMGTFET (dashed line) with $\Phi_{SMG} = 4.8$ eV, channel length 200 nm, $T_{si} = 10$ nm, $T_{ox} = 2$ nm, obtained by our model for $V_{DS} = -1.0$ V.

## V. CONCLUSIONS

In this paper, we have developed a compact analytical model for surface potential and drain current of a DMGTFET with equal lengths of tunneling and auxiliary gates and further extended it for a tunneling gate of smaller length. In our model we have included the effect of charge accumulation at the interface of the two gates and captured the change in tunneling volume with applied gate voltage. With this model, we have shown the step in the surface potential introduced by the DMG structure at the boundary of the two gates. We have also shown the occurrence of surface potential minimum in a DMGTFET with a smaller tunneling gate length. The model has been validated with accurate numerical simulations using the tunneling parameters extracted from reported experimental results. Using this model, we have demonstrated the potential advantages offered by a DMGTFET over a conventional SMGTFET as has already been demonstrated in literature [10]. We have shown how a DMGTFET offers a higher ON state current, a better $I_{ON}/I_{OFF}$ ratio and a lower $SS_{AVG}$ simultaneously. We have also shown how a DMGTFET solves the problem of delayed saturation exhibited by a SMGTFET. Hence, our model can be used for better understanding and prediction of the behavior of a DMGTFET.

## REFERENCES

[1] A. C. Seabaugh and Q. Zhang, "Low-voltage tunnel transistors for beyond CMOS logic," *Proc. IEEE*, vol. 98, no. 12, pp. 2095–2110, Dec. 2010.
[2] S. O. Koswatta, M. S. Lundstrom, and D. E. Nikonov, "Performance comparison between p-i-n tunneling transistors and conventional MOSFETs," *IEEE Trans. Electron Devices*, Vol. 56, No. 3, pp. 456–465, Mar. 2009.
[3] E.-H. Toh, G. H. Wang, G. Samudra and Y. C. Yeo, "Device physics and design of germanium tunneling field-effect transistor with source and drain engineering for low power and high performance applications", *J. Appl. Phys.,* Vol.103, No.10, pp. 104504 - 104504-5, 2008.
[4] M. J. Kumar and S. Janardhanan, "Doping-less Tunnel Field Effect Transistor: Design and Investigation", *IEEE Trans. on Electron Devices*, Vol.60, pp.3285-3290, October 2013.
[5] S. Saurabh and M. J. Kumar, "Estimation and Compensation of Process Induced Variations in Nanoscale Tunnel Field Effect Transistors (TFETs) for Improved Reliability," *IEEE Trans. on Device and Materials Reliability*, Vol.10, pp.390 - 395, September 2010.
[6] S. Saurabh and M. J. Kumar, "Impact of Strain on Drain Current and Threshold Voltage of Nanoscale Double Gate Tunnel Field Effect Transistor (TFET): Theoretical Investigation and Analysis," *Japanese Journal of Applied Physics*, Vol.48, paper no. 064503, June 2009.
[7] International Technology Roadmap for Semiconductor, http://www.itrs.net/.
[8] C. Shen, S.-L. Ong, C.-H. Heng, G. Samudra, and Y.-C. Yeo, "A Variational Approach to the Two-Dimensional Nonlinear Poisson's Equation for the Modeling of Tunneling Transistors", *IEEE Electron Device Lett.,* Vol.29, No.11, pp.1252-1255, 2008.
[9] K. Boucart and A. M. Ionescu, "A new definition of threshold voltage in Tunnel FETs," *Solid State Electron.*, Vol. 52, No. 9, pp. 1318–1323, Sep. 2008.
[10] S. Saurabh and M. J. Kumar, "Novel Attributes of a Dual Material Gate Nanoscale Tunnel Field-Effect Transistor", *IEEE Trans. Electron Devices,* Vol.58, No.2, pp.404-410, 2011.
[11] A. Zhang, J. Mei, L. Zhang, H. He, J. He and M. Chan, "Numerical study on dual material gate nanowire tunnel field-effect transistor", *IEEE International Conference on Electron Devices and Solid State Circuit (EDSSC)*, 2012, pp.1 - 5.
[12] N. Cui, R. Liang, and J. Xu, "Heteromaterial gate tunnel field effect transistor with lateral energy band profile modulation", *Applied Physics Letters,* Vol. 98, pp. 142105-142105-3, 2011.
[13] H. Lou, L. Zhang, Y. Zhu, X. Lin, S. Yang, J. He, and M. Chan, "A Junctionless Nanowire Transistor with a Dual-Material Gate", *IEEE Trans. Electron Devices*, Vol.59, No.7, pp.1829-1836, 2012.
[14] M. J. Lee and W. Y. Choi, "Effects of Device Geometry on Hetero-Gate-Dielectric Tunneling Field-Effect Transistors", *IEEE Electron Device Lett.,* Vol.33, No.10 pp.1459-1461, 2012.
[15] M.-L. Fan, V. P. Hu, Y.-N. Chen, P. Su and C.-T. Chuang, "Analysis of Single-Trap-Induced Random Telegraph Noise and its Interaction with Work Function Variation for Tunnel FET ", *IEEE Trans. Electron Devices,* Vol.60, No.6, pp.2038-2044, 2013.
[16] W. Vandenberghe, A. S. Verhulst, G. Groeseneken, B. Soree, and W. Magnus, "Analytical model for point and line tunneling in a tunnel field-effect transistor," in *Proc. Int. Conf. SISPAD*, 2008, pp. 137-140.
[17] W. G. Vandenberghe, A. S. Verhulst, G. Groeseneken, B. Soree and W. Magnus, "Analytical model for a tunnel field-effect transistor", *14th IEEE Mediterranean Electrotechnical Conference*, 2008, pp.923 - 928.
[18] A. S. Verhulst, D. Leoneli, R. Rooyackers, and G. Oroeseneken, "Drain voltage dependent analytical model of tunnel field-effect transistors," *J. Appl. Phys.,* Vol.110, pp.024510, 2011.
[19] V. Dobrovolsky and F. Sizov, "Analytical model of the thin-film silicon-on-insulator tunneling field effect transistor," *J. Appl. Phys.,* Vol.110, pp.114513, 2011.
[20] M. Lee and W. Choi, "Analytical model of single-gate silicon-on-insulator tunneling field-effect transistors," *Solid-State Electron.,* vo1.63, pp.110, 2011.
[21] L. Liu, D. Mohata, and S. Datta, "Scaling length theory of double-gate interband tunnel field-effect transistors," *IEEE Trans. Electron Devices*, Vo1.59, pp.902-908, 2012.
[22] J. Wan, C. L. Royer, A. Zaslavsky and S. Cristoloveanu, "A tunneling field effect transistor model combining interband tunneling with channel transport", *J. Appl. Phys.,* Vol.110, No.10, pp.104503 - 104503-7, 2011.
[23] M. G. Bardon, H. P. Neves, R. Puers and C. V. Hoof, "Pseudo-Two-Dimensional Model for Double-Gate Tunnel FETs Considering the Junctions Depletion Regions", *IEEE Trans. Electron Devices,* Vol.57, No.4, pp.827-834, 2010.
[24] E. O. Kane, "Theory of Tunneling", *J. Appl. Phys.,* Vol.32, No.1, pp.83-91, 1961.
[25] ATLAS Device Simulation Software, *Silvaco Int*., Santa Clara, CA, 2012.